# Mesoscopic entanglement of atomic ensembles through non-resonant stimulated Raman scattering

Wenhai Ji, Chunbai Wu, S.J. van Enk, M.G. Raymer

Oregon Center for Optics and Department of Physics,

University of Oregon, Eugene, 97403 OR, USA

Email: wji@uoregon.edu

(Dated: Mar. 20, 2007)

**Abstract:** We propose a scheme of generating and verifying mesoscopic-level entanglement between two atomic ensembles using non-resonant stimulated Raman scattering. Entanglement can be generated by direct detection or balanced homodyne detection of the Stokes fields from the two cells, after they interfere on a beam splitter. The entanglement of the collective atomic fields can be transferred to the anti-Stokes fields in a readout process. By measuring the operator moments of the anti-Stokes fields, we can verify the presence of entanglement. We model the effects of practical factors such as Stokes field detector quantum efficiency and additive thermal noise in the entanglement generating process, and anti-Stokes field losses in the entanglement verification process, and find achievable regimes in which entanglement can be verified at the levels of tens to hundreds of atomic excitations in the ensembles.

## 1. Introduction

Entanglement is a peculiar and fundamental feature of quantum mechanics. Entangled states are fundamental to quantum information, especially to quantum teleportation and dense coding of both discrete and continuous variables. Two physically separated quantum systems A and B are said to be entangled if their joint density operator cannot be represented as a convex sum of products of density operators for the two systems [1,2,3]. An entanglement criterion is needed to verify the entanglement. The criterion depends on the state to be measured and the variables to be entangled. For bipartite Gaussian-state entanglement, an inseparability criterion based on partial transposition (PT) was developed by Peres [1]. The PT approach was generalized by Shchukin and Vogel [4] to allow verifying multimode continuous-variable entanglement by the measurement of a set of operator moments. Miranowicz *et al*. [5] improved the Shchukin-Vogel continuous-variable approach by including more tests for the positivity of the matrix of moments. We propose a scheme of atomic-ensemble entanglement in the mesoscopic regime and analyze the entanglement with the Shchukin-Vogel [4] and Miranowicz *et al*. [5] approach.

Entanglement at the single-quantum level has been generated and verified between atomic ensembles by Chou *et al.* [6] and Matsukevich *et al*. [7], following the DLCZ (Duan, Lukin, Cirac, and Zoller) protocol [8]. In Chou *et al*. [6]'s scheme, quantum interference in the detection of single photons emitted by one of the two cesium ensembles projects the otherwise independent ensembles into an entangled state with one joint excitation shared by the two ensembles trapped in two remotely located magneto-optical traps (MOTs). They confirmed entanglement by mapping the state of the atoms to optical fields after a programmable delay and by measuring mutual coherences and photon statistics for these fields. They thereby determine a quantitative lower bound for the entanglement of the joint state of the ensembles. Matsukevich *et al.* [7] achieved entanglement by generating a correlated state of an atomic ensemble at magneto-optical trap (MOT) A and a Stokes field through spontaneous Raman scattering, transmitting the Stokes photon to atomic ensemble at MOT B, and converting the photon to an atomic-field excitation. With the help of control pulse, the entangled state was transferred from the atomic mode to photonic mode. By comparing polarization correlation of the subsequent optical field with Bell inequality, they can verify the entanglement. Julsgaard *et al.* [9] generated entanglement between collective spins of cesium ensembles in two cells by optical pumping them respectively with two opposite- circular polarized light fields. A probe pulse was sent through the two cells spatially in series and the atomic CSS entanglement was transferred to the probe optical field through the Faraday effect. By direct detection of Stokes operators of the probe field, Julsgaard verified a violation of the inequality derived by Duan *et al.* [2,3], a sufficient condition for entanglement.

In this paper, we propose a scheme for experimental verification of entanglement in the joint state of two atomic ensembles with large excitation number. Our mesoscopic entanglement generation scheme is based on non-resonant stimulated Raman scattering (SRS) and a beam splitter, proposed in [10] as an extension of the DLCZ protocol single-photon scheme [8]. Different from the direct detection and correlation measurement [6,7,9], our entanglement verification scheme requires statistical moments of annihilation and creation operators measurement. We find the limitations of the excitation number imposed by the performance of available photodetectors.

The quantum state stored in an atomic medium has a long decoherence time—both in a MOT and in a warm vapor cell, because the internal atomic coherence is insensitive to dephasing by Doppler broadening for the small frequency shift between the two lower levels. The internal state of the collective atomic pseudo-spin degree of freedom can be described using decomposition in terms of collective atomic ensemble modes. We refer to the collective atomic degrees of freedom as the atomic field. When all atoms or molecules are in a common state, called the ground state, we say that there are zero excitations in the atomic field. When there are excitations in the atomic field, these are delocalized and shared symmetrically by all atoms in the ensemble. This is analogous to a Dicke superradiant state [11]. For SRS, the atomic-field concept was used, for example, by Smithey *et al*. to quantify the level of coherent excitation per spatial atomic mode, following SRS and subsequent decay by collisional dephasing [12].

SRS is a third-order nonlinear optical process, illustrated in Figure 1. Stimulated Stokes field generation can be idealized as a coherent three-wave mixing process [13, 14]. Initially, all atoms occupy the same internal state (achieved by optical pumping). There are no excitations in the Stokes field or in the atomic field. After a non-resonant laser pump pulse (called the Stokes pump) passes through the vapor, Stokes photons and excited-state atoms are created in equal numbers. After a delay of hundreds or thousands of nanoseconds, a second non-resonant laser pulse (called the readout or anti-Stokes pump) is sent through the vapor in the same direction as the Stokes pump. By controlling the anti-Stokes pump pulse, the quantum state of the collective atomic field can be transferred to the outgoing anti-Stokes field [10,15] ideally in a unitary manner. The spatial phasing properties of the atomic ensemble during the storage period ensure that the generated anti-Stokes pulse exits as a well-collimated beam along the common axis [10,16-18]. Ideally, this leaves the atomic ensemble in its vacuum (ground) state.

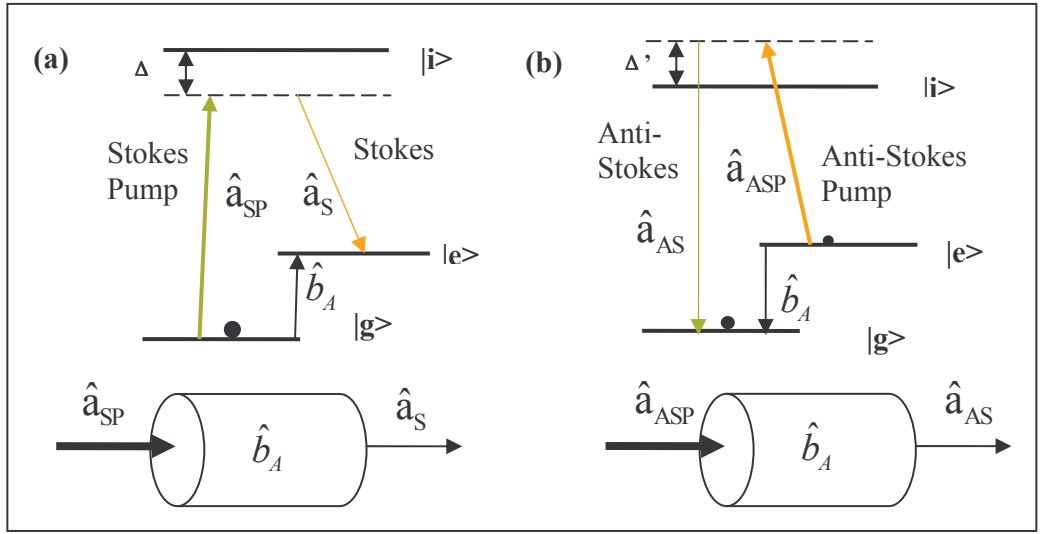

Fig.1. (Color online) (a): Three-level atomic system, with state g initially populated, is driven by detuned Stokes pump field, creating photons in Stokes field and excitations in atomic field. (b): Subsequently the anti-Stokes pump field drives the atoms to emit anti-Stokes photons and return to ground state g.

## 2. Conditional entanglement generation

Rigorous theoretical treatment of SRS process requires consideration of the spatial-temporal distribution of excitations of the fields. This generally results in multimode excitation of the collective atomic field and optical field. For these systems, the definition of modes is best done using spatial-temporal wave packets. Fortunately, theoretical calculation [10, 13-15] shows that in the absence of dispersion, single-mode excitation is a very good approximation. Wasilewski *et al.* [17] showed that even in the presence of weak dispersion, the excitation is strongly dominated by a single pair of spatial-temporal modes—one for the atomic field and one for the Stokes field. Therefore, in the following we assume that only a single wave-packet mode for each field is needed for an accurate description. Moreover, because the assumption of a single mode in the multimode situation is equivalent to a "local operation" of namely filtering from many local modes to a single local mode, our assumption can only lead to underestimating the entanglement generated [19].

In the Stokes scattering process, a two-mode squeezed state is developed between Stokes field and atomic field. Thus, Stokes photons and excited-state atoms are created in equal numbers. This creates, ideally, a pure entangled state of the atomic field and Stokes field,

$$| \Psi \rangle = \sum_{i=0}^{\infty} C_i \, | i \rangle_{Stokes} \otimes | i \rangle_{Atom} \qquad (1)$$

with

$$C_i = (\cosh \alpha)^{-1} (\tanh \alpha)^i \,,$$

where $\alpha$ is the squeezing parameter, determined by the Stokes pump pulse [13,14]. The fields' excitation number is denoted by $i$. The Stokes photon number obeys the geometric distribution, which can be reduced to an exponential distribution at large photon numbers [13, 14]. The measurement of the Stokes photon number also measures the atomic excitation number. Such a measurement allows one to predict the photon number of the anti-Stokes field in the later readout process.

We can generate, store and verify number-state entanglement between two atomic ensemble fields in two vapor cells A and B by the scheme shown in Figure 2. Following optical pumping (not shown), two Stokes pumps pass through the two cells, creating excitations in the atomic and Stokes fields. The Stokes fields interfere at a 50:50 beam splitter (BS), and the resultant fields are measured by ideal number-resolving detectors. (We consider imperfect detectors later in the paper). Knowledge of the measured Stokes photon numbers provides one with an entangled-state description of the atomic fields in the two cells. The entanglement can be stored in the cells for some time, can be transferred to anti-Stokes fields later, and can subsequently be verified by measuring the anti-Stokes fields.

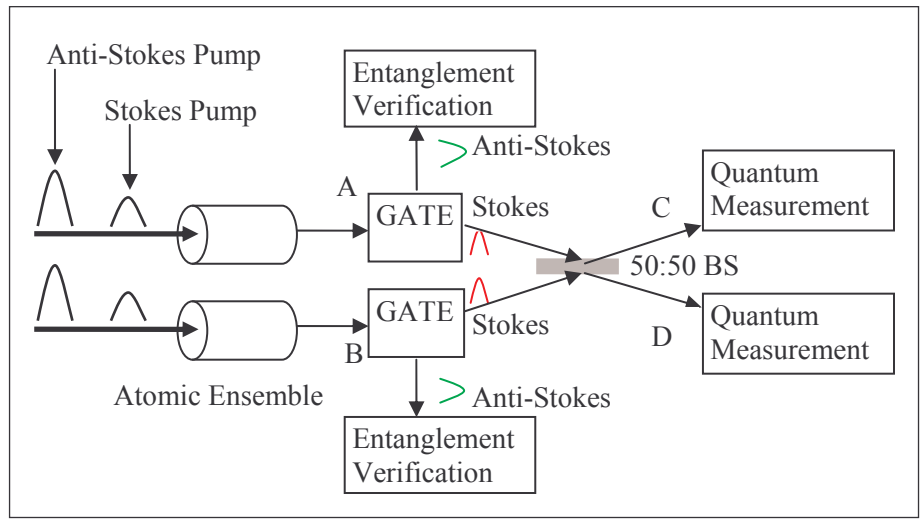

Fig.2: (Color online) Conditional entanglement generation and storage scheme

The state of the Stokes fields and atomic fields for cells A and B before the BS is the direct product state of the two individual cells,

$$|\Psi\rangle_{A,B} = |\Psi\rangle_A \otimes |\Psi\rangle_B \qquad (2)$$

where $|\Psi\rangle_A$, $|\Psi\rangle_B$ are quantum states (of the form of Eq. (1)) of cells A and B. After the 50:50 BS, the resultant Stokes fields are detected by assumed perfect photon detectors C and D, which register $n$ and $m$ photons. Conditioned on a group of detection outcome pairs with the same ($n$, $m$) values, we can infer the conditioned atomic state as

$$|\Phi\rangle_{A,B}^{Atom} = (\cosh\alpha)^{-1}(\cosh\beta)^{-1}\sum_{i=0}^{\infty}\sum_{j=0}^{\infty}(\tanh\alpha)^i(\tanh\beta)^j B_i^{n,m}|i,j\rangle_{Atom}, \qquad (3)$$

where

$$B_i^{n,m} = \langle n,m|\hat{B}^+|i,j\rangle = \delta_{i+j-n-m}\sum_{k=0}^{i\,\text{or}\,n}(-1)^{i-k}\left(\frac{1}{\sqrt{2}}\right)^{n+m}\binom{n}{k}\binom{m}{i-k}\sqrt{\frac{i!(n+m-i)!}{n!m!}} \quad.$$

In Eq. (3), $\alpha$, $\beta$ are the squeezing parameters of cells A and B, and $i$ and $j$ are field excitation numbers of cells A and B. The conservation relation $i + j = n + m$ holds. Derivation of the BS coefficient $B_i^{n,m}$ is a standard procedure and can be found in [20-22].

A perfect readout of atomic field state can be obtained in principle by controlling the anti-Stokes pump pulse [10]. In the perfect readout process, the state of anti-Stokes field is the same as that of atomic field. In Stokes scattering process, the atomic field and Stokes field are created in equal excitation numbers. Therefore, a strong positive correlation between Stokes photons and anti-Stokes photons is expected and observed [23-25]. The joint state of the anti-Stokes fields is,

$$|\Phi\rangle_{A,B}^{AS} = (\cosh\alpha)^{-1}(\cosh\beta)^{-1}\sum_{i=0}^{n+m}(\tanh\alpha)^i(\tanh\beta)^{n+m-i}\,B_n^{i,n+m-i}\,|i\rangle_A^{AS}\otimes|n+m-i\rangle_B^{AS} \quad (4)$$

To normalize the state, we divide Eq. (4) by the probability amplitude of detecting $i$ and $n+m-i$ Stokes photons. If the squeezing parameters are equal ( $\alpha = \beta$ ), this gives an especially simple result for the state of the anti-Stokes fields, which is independent of the values of the squeezing parameters,

$$|\Phi\rangle_{A,B}^{AS} = \sum_{i=0}^{n+m} B_n^{i,n+m-i}\,|i\rangle_A\otimes|n+m-i\rangle_B \quad (5)$$

This states, simply, that if $n$ and $m$ Stokes photons are detected, the total number of inferred atomic excitations $n+m$ is shared symmetrically by the two ensembles, according to a BS transformation. Mapping the atomic ensemble state to a state of the anti-Stokes fields is a local operation and cannot increase the average amount of entanglement. Hence, the entanglement inferred between the anti-Stokes fields is not more than that between the two ensembles.

An effective model is sometimes helpful. If we treat the measured Stokes numbers ( $n$ , $m$ ) as an "effective BS input", then the joint state of the corresponding BS output has the same expression as in Eq. (5), for the anti-Stokes fields, up to phase factors. Thereby, the backward inference of the conditioned state following measurement has the same mathematical form as the forward prediction of the output state for a 50:50 BS transformation.

## 3. Characterization of the entangled state

The anti-Stokes state of Eq. (5) is an entangled state, and our task is to verify this experimentally. As part of the characterization of the entangled state, we first study the conditional correlation between the two anti-Stokes fields,

$$C = \frac{\langle n_A n_B \rangle - \langle n_A \rangle \langle n_B \rangle}{\sqrt{\langle n_A^2 \rangle - \langle n_A \rangle^2} \sqrt{\langle n_B^2 \rangle - \langle n_B \rangle^2}} \tag{6}$$

where $n_A$, $n_B$ are measured photon number of the anti-Stokes fields from cells A and B. In the Heisenberg picture, the annihilation operators of anti-Stokes fields $\hat{a}$, $\hat{b}$ are related to those of the Stokes fields $\hat{c}$, $\hat{d}$ by $\hat{a} = 2^{-1/2}(\hat{d} + \hat{c})$, $\hat{b} = 2^{-1/2}(\hat{d} - \hat{c})$. This follows from our recognition that Eq. (5) is simply a 50:50 BS transformation. With the state $|n\rangle_C |m\rangle_D = |n, m\rangle$ inferred from the detected Stokes fields, the quantities in Eq. (6) are:

$$\langle n_A n_B \rangle = \langle n, m | a^\dagger a b^\dagger b | n, m \rangle = \frac{1}{4}(n^2 + m^2 - n - m)$$

$$\langle n_A \rangle = \langle n_B \rangle = \frac{1}{2}(n + m), \ \langle n_A^2 \rangle = \langle n_B^2 \rangle = \frac{1}{4}(n^2 + m^2 + 4nm + n + m) \tag{7}$$

These quantities can also be obtained in the Schrödinger picture, giving identical results. Putting these quantities into the conditional correlation formula gives $C = -1$. The perfect anti-correlation agrees with our expectation based on our knowledge of the total number of anti-Stokes photons, conditioned on the same measured photon number pair ($n$, $m$) of the resultant Stokes fields.

The Shchukin-Vogel CV extension approach [4,5] for entanglement verification relies on the properties of certain combinations of moments under the operation of density matrix partial transposition. The approach starts from tensor product classes of operators,

$$\hat{f} = (\hat{a}^{\dagger k_1^{(1)}} \hat{a}_2^{k_2^{(1)}},...,\hat{a}^{\dagger k_1^{(d_A)}} \hat{a}_2^{k_2^{(d_A)}}) \otimes (\hat{b}^{\dagger l_1^{(1)}} \hat{b}_2^{l_2^{(1)}},...,\hat{b}^{\dagger l_1^{(d_B)}} \hat{b}_2^{l_2^{(d_B)}})$$
$$= (...,\hat{a}^{\dagger k_1^{(i)}} \hat{a}_2^{k_2^{(i)}} b^{\dagger l_1^{(j)}} \hat{b}_2^{l_2^{(j)}}, ...) = (...,\hat{f}^{k^{(i)},l^{(j)}}, ...) \tag{8}$$

where $k^{(i)}$ and $l^{(j)}$ are two-dimensional vectors. The corresponding moment matrix $M_{\hat{f}}(\hat{\rho})$ for the density operator $\hat{\rho}$ and the moment matrix $M_{\hat{f}}(\hat{\rho}^\Gamma)$ for the partially transposed (on subsystem B only) density operator $\hat{\rho}^\Gamma$ are related by,

$$M_{\hat{f}}(\hat{\rho}) \equiv \sum_{kl,k'l'} M_{kl,k'l'}(\hat{\rho}) \,|kl\rangle\langle k'l'| = \sum_{kl,k'l'} \langle \hat{f}^{k,l} \cdot \hat{f}^{k',l'}\rangle \,|kl\rangle\langle k'l'|$$
$$M_{\hat{f}}(\hat{\rho}^\Gamma) \equiv \sum_{kl,k'l'} M_{kl,k'l'}(\hat{\rho}) \,|kl'\rangle\langle k'l| = \sum_{kl,k'l'} \langle \hat{f}^{k,l} \cdot \hat{f}^{k',l'}\rangle \,|kl'\rangle\langle k'l| \tag{9}$$

For any separable density matrix, the determinants of $M_{\hat{f}}(\hat{\rho})$ and $M_{\hat{f}}(\hat{\rho}^\Gamma)$ are nonnegative and all of their eigenvalues are nonnegative. If we can find a $\hat{f}$ such that the measured moment matrix (or any of its submatrices) for $\hat{\rho}^\Gamma$ has negative eigenvalues or determinant(s), then entanglement of that state is verified. These conditions are only sufficient, not necessary, conditions for the presence of entanglement.

The SV criterion [4] and its extension [5] are given in terms of creation and annihilation operator moments, which are measurable using quantum state tomography [26,27], phase randomized balanced homodyne detection (BHD) [28,29], or balanced homodyne correlation (BHC) [30,31]. We can measure the quadrature distribution versus the relative phase $\Phi$ between signal and local oscillator (LO) in the standard

BHD measurement (Figure 3(a)). The normally-ordered single-mode field moments, $\langle \hat{a}^{+k} \hat{a}^{l} \rangle$, could be obtained directly from the quadrature distribution [28,29],

$$\langle \hat{a}^{\dagger k} \hat{a}^{l} \rangle = \left[ \pi \sqrt{2^{k+l}} \binom{k+l}{k} \right]^{-1} \langle H_{k+l}(q_{a,\Phi}) e^{i(k-l)\Phi} \rangle_{\Phi} \qquad (10)$$

where $H_{k+l}(\bullet)$ is the $(k+l)$-th order Hermite polynomial function and $\langle \bullet \rangle_{\Phi}$ represents the corresponding quantity uniformly averaged over all possible values of $\Phi$. Although more complicated than BHD, the BHC measurement (Figure 3(b)) is claimed to have less noise and more direct results for the moments. For example, if d detectors and $2^d+1$ beam splitters are used, we can measure any moment $\langle \hat{a}^{\dagger k} \hat{a}^{l} \rangle$ with $k$, $l = 0, 1, 2, \cdots, 2^{d}$ [30,31]. The general form of joint moments in Eq. (9) can be converted to normal ordered form $\langle \hat{a}^{+k} \hat{b}^{+l} \hat{a}^{i} \hat{b}^{j} \rangle$ and thus can be obtained by making joint BHD or BHC measurement of fields A and B. Some lower order moments like $\langle \hat{a}^{\dagger} \hat{b} \rangle$ or $\langle (\hat{a}^{\dagger} \hat{b})^2 \rangle$ can also be obtained through eight-port detection [32], in which no LO is required.

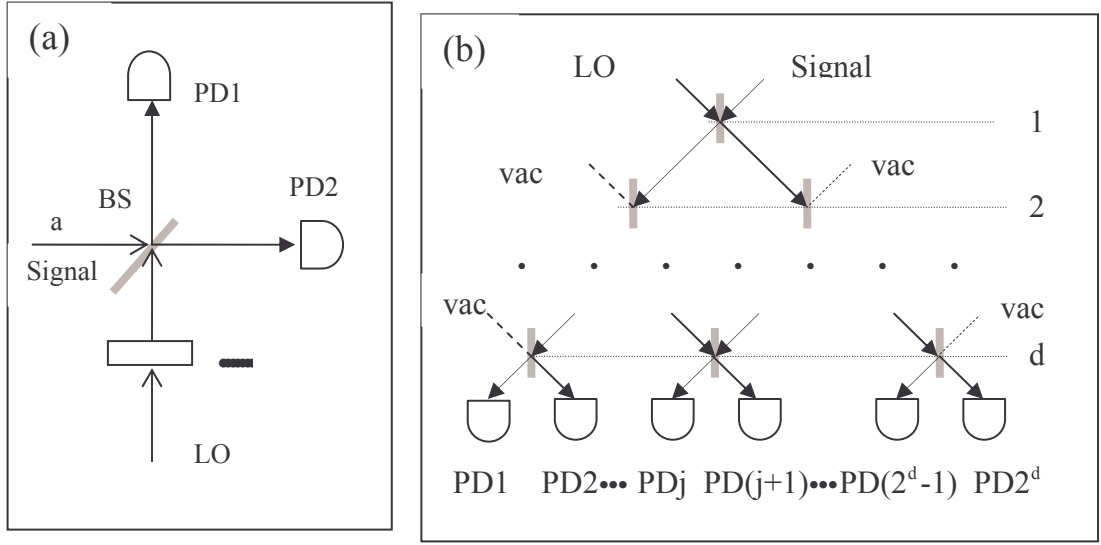

**Fig.3 (a): Balanced homodyne detection measurement of quadratures. (b): Balanced homodyne correlation measurement of matrix moment.**

In order for the moment $\langle \hat{a}^{\dagger k} \hat{b}^{\dagger l} \hat{a}^{i} \hat{b}^{j} \rangle$ to be non-zero, there are some constraints for $k$ , $l$ , $i$ , $j$ , which arise from the specific entangled state form of Eq. (5). Conservation of photon number in the BS transformation requires that $k + l = i + j$ . For a symmetric effective BS "input" (that is, Stokes field measurement result, $n = m$ ), $k + l = i + j$ must be an even number to keep the moment non-zero. Thus in our theoretical treatment, we simply choose $k = i$ and $l = j$ to be even, and the tensor function $\hat{f} = (1, \hat{a}^2 \hat{b}^2, \hat{a}^4 \hat{b}^4, \hat{a}^6 \hat{b}^6, \bullet\bullet\bullet, \hat{a}^{2k} \hat{b}^{2k} \bullet\bullet\bullet)$ . In the Heisenberg picture, the PT matrix moments are,

$$
\begin{aligned}
\langle \hat{a}^{\dagger k} \hat{b}^{\dagger l} \hat{a}^{l} \hat{b}^{k} \rangle &= \frac{1}{2^{k+l}} \sum_{h=0}^{k+l} \sum_{r,i=0,0}^{k,l} (-1)^{r+i} \binom{k}{r} \binom{l}{h-r} \binom{l}{i} \binom{k}{h-i} \langle \hat{c}^{\dagger h} \hat{c}^{h} \hat{d}^{\dagger (k+l-h)} \hat{d}^{(k+l-h)} \rangle \\
&= \sum_{h=0}^{k+l} g(h) \frac{n!}{(n-h)!} \frac{m!}{(m-(k+l-h))!}
\end{aligned}
\tag{11}
$$

where

$$
g(h) = \frac{1}{2^{k+l}} \sum_{r,i=0,0}^{k,l} (-1)^{r+i} \binom{k}{r} \binom{l}{h-r} \binom{l}{i} \binom{k}{h-i}
$$

An equivalent formula can also be obtained in the Schrödinger picture with the number-state representation, and it yields identical results as Eq. (11). With the number-state representation and the Schrödinger picture approach, we can predict the joint state of anti-Stokes fields from cells A and B. On the other hand, if only moments are needed, the Heisenberg picture approach is sufficient. Because this approach doesn't involve the complicated BS coefficients, it is more convenient from a computational point of view. In addition, it gives analytical expressions for the matrix determinant. Further advantages will be more apparent when practical factors are considered in the model.

The more terms that are included in the tensor function $\hat{f}$, the more powerful it can be for detecting entanglement. But in experiment and in simulation, it is unrealistic to take $k$ to be too large. We introduce a new notation for tensor functions: $\hat{f}_{i,j,\cdots,k}$ denotes a tensor function that includes $i$-th, $j$-th, $\cdots k$-th terms of $\hat{f}$. For example, $\hat{f}_{1,2,3,4} = (1,\ \hat{a}^2\hat{b}^2, \hat{a}^4\hat{b}^4, \hat{a}^6\hat{b}^6)$. For simplicity, let us initially assume there are equal detected Stokes photon numbers ($n = m$). Then we find that for each $\hat{f}_{i,j,\cdots,k}$, there is a maximum "input" photon number ($MaxN$) for which the entanglement can be verified. If the photon number $n = m$ is below $MaxN$, the PT matrix has negative eigenvalues. But the determinant is not necessary negative, because the PT matrix may have even numbers of negative eigenvalues. For example, for the tensor function $\hat{f}_{1,2,3,4}$, the PT matrix in the $n = m = 1\cdots16$ region has two negative eigenvalues and a positive determinant. If $n$ is above $MaxN$, the PT matrix no longer has negative eigenvalues and its determinant is always positive, even though the state is in fact entangled. For this reason we define the "test power" of the tensor function to be $MaxN$.

The relation of $MaxN$ versus the tensor function for $n = m$ is in shown in Table 1. It shows that $\hat{f}_{1,2} = (1, \hat{a}^2 \hat{b}^2)$ allows verification of entanglement up to $n = 13$, while $\hat{f}_{1,2,3,4}$ allows verification of entanglement up to $n = 444$ (assuming ideal detectors).

| Tensor Function | $\hat{f}_{1,2}$ | $\hat{f}_{1,3}$ | $\hat{f}_{1,4}$ | $\hat{f}_{1,5}$ | $\hat{f}_{2,3}$ | $\hat{f}_{1,2,3}$ | $\hat{f}_{2,3,4}$ | $\hat{f}_{1,2,3,4}$ |
|---|---|---|---|---|---|---|---|---|
| $MaxN$ | 13 | 28 | 48 | 72 | 91 | 114 | 403 | 444 |

**Table 1.** $MaxN$ **versus tensor function for** $n = m$

If $n = m$, we can get the determinant for $\hat{f}_{1,2}$

$$Det\left(M_{\hat{f}}(\hat{\rho}^{\Gamma})\right) = -\frac{3}{4}n + \frac{11}{8}n^2 - \frac{7}{4}n^3 + \frac{1}{8}n^4 . \tag{12}$$

Further checking determinants for other tensor function such as $\hat{f}_{1,2,3}$ and $\hat{f}_{1,2,3,4}$ with number of element in the tensor function $r = 3,\ 4$ respectively, we find that the determinant has a polynomial expression in the photon number, with highest degree equal to $d = 4 \times (i - 1 + j - 1 + \cdots + k - 1)$ and corresponding coefficient of $2^{r-d-1}$. All determinants eventually become positive as the photon number increases. This qualitatively explains the existence of the $MaxN$ for each tensor function in Table 1.

Now, we relax the assumption that $n = m$. Figure 4(a) and (b) are the contour plots of regions having detectable entanglement when we use tensor function $\hat{f}_{1,2,3}$ and $\hat{f}_{1,2,3,4}$ in the detected photon number pairs ($n$, $m$) of Stokes fields. The shaded regions are entanglement detectable regions. The darkness of the regions represents the probability of observing such pairs for fixed equal squeezing parameter $\alpha = \beta$ in the Stokes scattering process. The Stokes photon number has an exponential distribution, with lower photons having higher observable probability, as indicated by the darker color in Figure 4(a) and (b). The diagonal direction in the plots represents a symmetrical situation where $n = m$. We define a priori entanglement detection probability as the portion of entangled detectable regions among all possible regions weighted by observable probability, given a squeezing parameter $\alpha$. Figure 4(c) is plot for priori entanglement detectable probability versus $\alpha$.

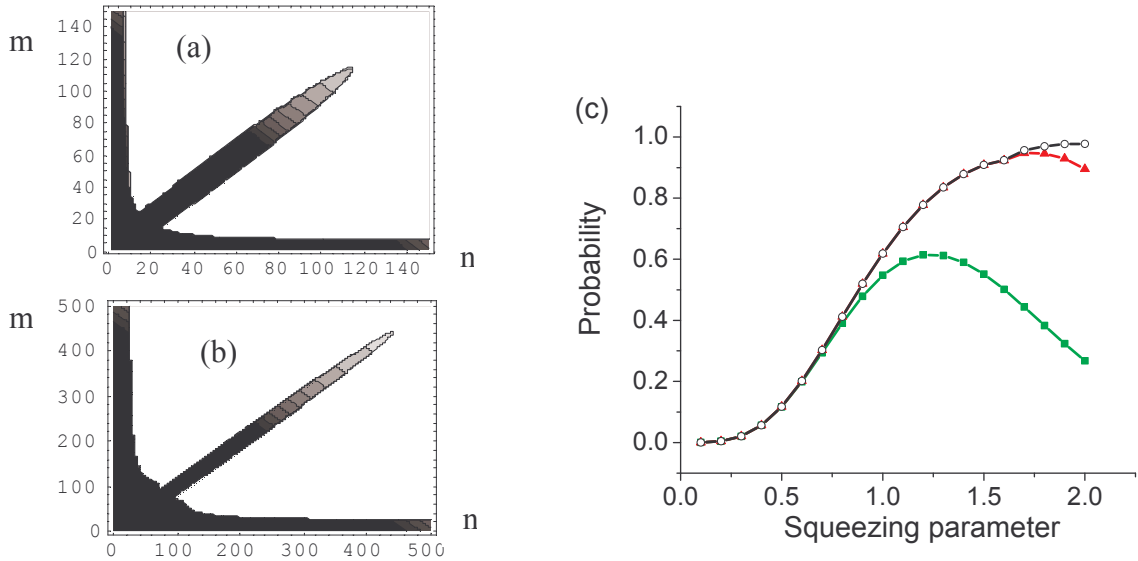

**Fig.4 (a) and (b):** The contour plots of entanglement detectable regions in the two detected photoelectron number ($n$, $m$) pairs from detection of Stokes fields when we use tensor function $\hat{f}_{1,2,3}$ (a) and $\hat{f}_{1,2,3,4}$ (b). The shaded regions are regions of detectable entanglement. The darkness of the regions represents the probability of observing such pairs for fixed equal squeezing parameter. (Color online) (c): A priori entanglement detectable probability versus squeezing parameter $\alpha = \beta$ for the tensor function $\hat{f}_{1,2}$ (solid square), $\hat{f}_{1,2,3}$ (solid triangle) and $\hat{f}_{1,2,3,4}$ (empty circle).

From Table 1 and Fig. 4(a) and (b), we observe that the test power of a tensor function will increase as extra terms are included in the tensor function. For tensor function with same number of terms, the tensor function including higher order terms has higher test power. The test power of $\hat{f}_{1,2,3}$ is higher than the test power of any its sub tensor function: $\hat{f}_{1,2}$ , $\hat{f}_{1,3}$ and $\hat{f}_{2,3}$ . The existence of non-shaded regions means that not every entangled state can be verified by either $\hat{f}_{1,2,3}$ or $\hat{f}_{1,2,3,4}$ . For a state of the form Eq. (5) with any specified values of ( $n$ , $m$ ), there exists such a tensor function with higher orders and more terms that can be used to verify the entanglement.

## 4. Degradation of entanglement by imperfect Stokes field detector and anti-Stokes field losses

The above calculations were made assuming an ideal situation: single modes of Stokes field and anti-Stokes field, noiseless and perfect Stokes photon detection, and lossless anti-Stokes field during readout and BHC moment measurement process. We need to consider the effects of the real world factors. In this situation, the measured "photon numbers" of Stokes field and anti-Stokes field actually refer to the photoelectron numbers resulting from the detection performed on the corresponding fields. Because of these factors, the photoelectron numbers don't necessarily equal the corresponding photon numbers.

Generally, the effect of noise on the entanglement generated in different stages of the process is not the same. Noise during the entanglement generation process always degrades the entanglement. Known noise during the verification process, in contrast, can be taken into account in the analysis of the entanglement, and so does not necessarily weaken the verification [19].

*4.1 The predicted anti-Stokes state under different practical factors*

In the Schrödinger picture, we can predict the joint state of the anti-Stokes field, conditioned on the measured Stokes photoelectron number. To predict the joint state under different practical situations, we need to model practical factors such as Stokes field detector quantum efficiency and noise, and anti-Stokes field losses. We compute the photon number probability distribution in the number-state representation. In the ideal situation, the anti-Stokes state is described by Eq. (5). The marginal photoelectron probability distribution of anti-Stokes field from one cell is,

$$p_i = |B_i^{n,m}|^2 . \qquad (13)$$

We first consider the uncertainty caused by the Stokes field detector efficiency and noise. Since there is no way to obtain the phase information about the Stokes state, we treat it as a mixed state. Then the anti-Stokes field has marginal photon number distribution

$$p_r = \sum_{i,j=0,0}^{\infty,\infty} p_C(i \,|\, n) p_D(j \,|\, m) |B_r^{i,j}|^2 . \qquad (14)$$

where $p_C(i \,|\, n)$ ( $p_D(j \,|\, m)$ ) is the conditional probability of having $i$ ( $j$ ) Stokes photons when $n$ ( $m$ ) photoelectrons are recorded by Stokes field detector C(D). Here we consider the detector quantum efficiency and additive noise separately. When $n$ (m) photons strike on a detector with quantum efficiency $\eta_C$ ( $\eta_D$ ), the photoelectron number has a binomial distribution. We assume a uniform Stokes photon number prior distribution. The assumption is valid if the variance of photoelectron binomial distribution is much smaller than that of Stokes photon number exponential distribution. Through Bayesian analysis, we get the probability of having $i$ ( $j$ ) input photons when $n$ ( $m$ ) photoelectrons are recorded, which is a negative binomial distribution,

$$
\begin{aligned}
p_C(i\,|\,n) &= \binom{i}{n}\eta_C^{n+1}(1-\eta_C)^{i-n} \\
p_D(j\,|\,m) &= \binom{j}{m}\eta_D^{m+1}(1-\eta_D)^{j-m}
\end{aligned}
\tag{15}
$$

Detector electronic thermal noise also leads to uncertainty of the actual Stokes photon number. The probability of having $i$ ( $j$ ) Stokes photons when $n$ ( $m$ ) photoelectrons detected is approximately a Gaussian distribution,

$$
\begin{aligned}
p_C(i\,|\,n) &= \frac{1}{\sqrt{2\pi}\sigma_C}\exp(-\frac{(i-n)^2}{2\sigma_C^2}) \\
p_D(j\,|\,m) &= \frac{1}{\sqrt{2\pi}\sigma_D}\exp(-\frac{(j-m)^2}{2\sigma_D^2})
\end{aligned}
\tag{16}
$$

where $\sigma_C$ and $\sigma_D$ are standard deviations of the electronic additive noise of Stokes detectors C and D and they given as equivalent noise photon numbers.

Losses in the anti-Stokes field can arise from the imperfect readout from the ensemble or from the propagation paths. Additional losses occurs during optical detection of the anti-Stokes fields, such as mode-mismatch between anti-Stokes and LO, and the detector quantum efficiency. We model all these together by an effective BS interaction with effective transmission $\eta_A$ ( $\eta_B$ ),

$$\eta_A = \eta_A^{Read} \times \eta_A^{Detect}$$
$$\eta_B = \eta_B^{Read} \times \eta_B^{Detect}$$

(17)

where $\eta_A^{Read}, \eta_A^{Detect}$ ($\eta_B^{Read}, \eta_B^{Detect}$) are the transmission efficiency of anti-Stokes field from cell A (B) in readout and moments measuring process respectively. We treat the atomic field as one input of a BS, and detected anti-Stokes field as the transmitted BS output. The other input of the BS is vacuum. The expression of the joint anti-Stokes state is,

$$\Psi_{A,B} = \sum_{i,r,s=0,0,0}^{n+m,i,n+m-i} B_i^{n,m} b_{r,s}^{i,n+m-i} \mid r \rangle_A \mid s \rangle_B \mid i-r \rangle_A^{Loss} \mid n+m-i-s \rangle_B^{Loss}$$
$$\text{where } b_{r,s}^{i,n+m-i} = \sqrt{\binom{i}{r}\binom{n+m-i}{s}\eta_A^r(1-\eta_A)^{i-r}\eta_B^s(1-\eta_B)^{n+m-i-s}}$$

(18)

where $i$ and $n+m-i$ are anti-Stokes photoelectron numbers without any losses, $r$ and $s$ are photoelectron number resulting from detection of the anti-Stokes fields. Then the anti-Stokes photoelectron number distribution is,

$$p_r = (\sum_{i,\ s=0,0}^{n+m,\ n+m-i} B_i^{n,m} b_{r,s}^{i,n+m-i})^2$$

(19)

With these probability formulas, the photoelectron probability of one anti-Stokes field versus photon number, when the detected Stokes photoelectron numbers are $n = m = 10$, is plotted in Figure 5. For simplicity, we consider these practical factors separately. When one factor is modeled, the other process is assumed to be ideal. In the all-ideal situation, the photoelectron probability distribution curve has the oscillating and parabolic shape, which reflects the Bosonian nature of photons. For anti-Stokes field measurement with effective transmission efficiency $\eta_A = \eta_B = 50\%$, the anti-Stokes field has less average photon number than in the ideal situation. For uncertainty caused by Stokes field detector efficiency $\eta_C = \eta_D = 90\%$ and detector thermal noise standard deviation $\sigma_C = \sigma_D = 2$, the oscillating feature and parabolic shape of the probability distribution curves are still observable, although they are not as apparent as in the ideal situation.

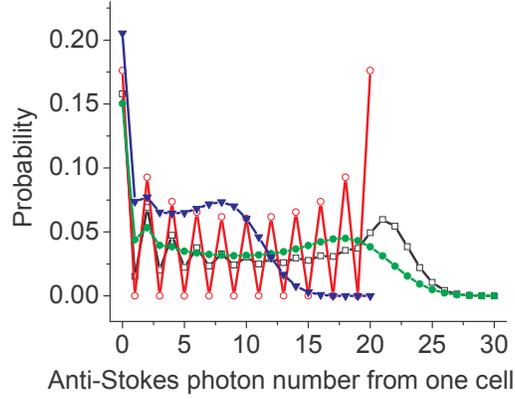

**Fig.5: (Color online) The predicted anti-Stokes state photoelectron-number probability when 10 photoelectrons are recorded in each of the Stokes field detector C and D for ideal situation (empty circle), for anti-Stokes field effective transmission efficiency $\eta_A = \eta_B = 50\%$ (solid triangle), with uncertainty caused by Stokes field detector quantum efficiency $\eta_C = \eta_D = 90\%$ (empty square) and detector thermal noise $\sigma_C = \sigma_D = 2$ (solid circle).**

*4.2 Effect of uncertain Stokes state caused by detectors with detection inefficiency and noise*

Because of the imperfect detection caused by detector inefficiency and additive electronic noise, the exact Stokes photon number incident on the detectors is not known from the detected photoelectron number. The Stokes state can only be inferred based on photoelectrons and their probability distribution. Now we assume that we have effective transmission efficiency of 100% for anti-Stokes field from cells A and B. Then the quantities in the conditional correlation equation (Eq. (6)) are,

$$\langle n_A n_B \rangle = \frac{1}{4} \sum_{i,j} p_C(i \mid n) p_D(j \mid m)(i^2 + j^2 - i - j), \ \langle n_A \rangle = \langle n_B \rangle = \frac{1}{2} \sum_{i,j} p_C(i \mid n) p_D(j \mid m)(i + j)$$

$$\langle n_A^2 \rangle = \frac{1}{4} \sum_{i,j} p_C(i \mid n) p_D(j \mid m)(i^2 + j^2 + i + j + 4ij)$$

$$(20)$$

If $p_C(i \mid n)$, $p_D(j \mid m)$ are identical probability distributions, it is straightforward to show that the conditional correlation C is still -1, that is, the inferred correlation is perfect.

The elements of the PT matrix for this mixed state in the Heisenberg picture are,

$$\langle \hat{a}^{+k} \hat{b}^{+l} \hat{a}^l \hat{b}^k \rangle = \sum_{i,j=n,m}^{\infty,\infty} p_C(i \mid n) p_D(j \mid m) \sum_{h=0}^{k+l} g(h) \frac{i!}{(i-h)!} \frac{j!}{(j-(k+l-h))!} \qquad (21)$$

We still choose $n = m$ and assume that Stokes field detectors C and D have identical noise performance, $\eta_C = \eta_D$ and $\sigma_C = \sigma_D = \sigma$. Figure 6(a) shows the *MaxN* versus Stokes field detector efficiency for tensor function $\hat{f}_{1,2,3}$ and $\hat{f}_{1,2,3,4}$. The *MaxN* drops linearly as detector efficiency decreases. Figure 6(b) shows the range of photoelectrons with corresponding anti-Stokes field entanglement verified, versus the standard deviation of detector additive thermal noise. For each tensor function and each $\sigma$, there exists *MaxN* and *MinN*. That is, if $n = m$ falls within the *MinN* to *MaxN* region, the PT matrix has negative eigenvalue, otherwise it is positive. The figure shows that for a realistic detector with $\sigma$ around 6, photoelectron number for $\hat{f}_{1,2,3,4}$ verified entanglement is in the mesoscopic range of about 50 to 400. The curve becomes closed at a point slightly below $MaxN\big/2$ when $\sigma$ reaches some critical value. The PT matrix has positive determinants and no negative eigenvalues, if $\sigma$ >10.545 for $\hat{f}_{1,2,3,4}$, or $\sigma$ >5.35 for $\hat{f}_{1,2,3}$.

The figure shows that the entanglement detectable region shrinks as $\sigma$ increases and the electronic noise has an adverse effect on the entanglement verification. So the uncertainty in the entanglement generating process caused by Stokes field detector degrades the created entanglement. Still, present detector technology should be sufficient for achieving and verifying entanglement in this mesoscopic regime.

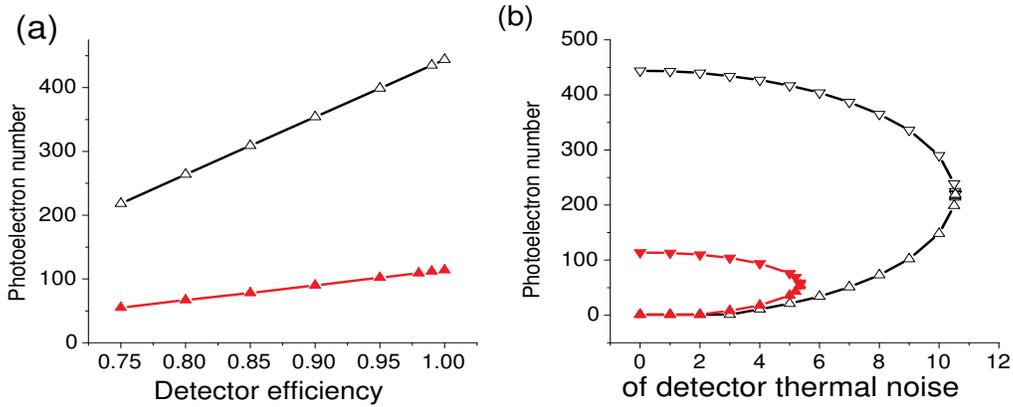

Fig.6 (a): (Color online) The entanglement test power of tensor function $\hat{f}_{1,2,3}$ (solid triangle) and

$\hat{f}_{1,2,3,4}$ (empty triangle) for joint state of anti-Stokes fields from cells A and B versus the Stokes field

detector efficiency. (b): The test power of tensor function $\hat{f}_{1,2,3}$ (solid triangle) and $\hat{f}_{1,2,3,4}$ (empty triangle) versus $\sigma$, the standard deviation of Stokes field detector total noise. For each $\sigma$, there is a maximum photoelectron number $MaxN$ (top part of the curve) and minimum photoelectron number $MinN$ (lower part of the curve).

*4.3. Effect of anti-Stokes field losses during the verification process*

As mentioned in the previous section, anti-Stokes field losses during the verification process can be modeled by an effective BS interaction. In the Heisenberg picture, the field operators are

$$\hat{a} = \sqrt{\eta_A}\hat{a}_a + \sqrt{1-\eta_A}\hat{V}, \quad \hat{b} = \sqrt{\eta_B}\hat{b}_a + \sqrt{1-\eta_B}\hat{U} \qquad (22)$$

where $\hat{a}$, $\hat{b}$ are annihilation operators of the effective detected anti-Stokes fields, $\hat{a}_a$, $\hat{b}_a$ are annihilation operators of the atomic fields, $\hat{U}$, $\hat{V}$ are annihilation operators of vacuum fields, and $\eta_A$, $\eta_B$ are effective transmission efficiencies of anti-Stokes field from cells A and B. We assume that $\eta_A = \eta_B = \eta$ and the Stokes field detectors are ideal, 100% quantum efficiency and no thermal noise.

The quantities in the conditional correlation equation (Eq. (6)) for the measured anti-Stokes photoelectron numbers are,

$$\langle n_A n_B \rangle = \frac{1}{4}\eta^2(n^2 + m^2 - n - m), \ \langle n_A \rangle = \langle n_B \rangle = \frac{1}{2}\eta(n+m)$$

$$\langle n_A^2 \rangle = \frac{1}{4}\eta^2(n^2 + m^2 + 4nm + n + m) + \frac{1}{2}\eta(1-\eta)(n+m) \qquad (23)$$

$$C = -\frac{1}{1 + \dfrac{2(1-\eta)}{\eta}\dfrac{n+m}{n+m+2nm}}$$

The conditional correlation C can still approach -1, if we condition on bigger Stokes photoelectron number $n$, $m$ or increase the effective transmission efficiency $\eta$, so that the term caused by vacuum fluctuation plays a smaller roll in the denominator. This occurs for $\eta \gg (n+m)/nm$.

The moments of the PT matrix are,

$$\langle \hat{a}^{\dagger k} \hat{b}^{\dagger l} \hat{a}^{l} \hat{b}^{k} \rangle = (\sqrt{\eta_A \eta_B})^{k+l} \sum_{h=0}^{k+l} g(h) \frac{n!}{(n-h)!} \frac{m!}{(m-(k+l-h))!} \qquad (24)$$

where $g(h)$ is the same as in Eq. (11). The result shows that for a practical anti-Stokes state with density operator $\hat{\rho}_\eta$, the determinant of the PT matrix for tensor function $\hat{f}_{i,j,\cdots,k}$ with $r$ elements is related to the matrix determinant for the ideal state with density operator $\hat{\rho}$ by

$$Det(M(\hat{\rho}_\eta^\Gamma)) = (\eta_A \eta_B)^{2(i+j+\cdots+k-r)} Det(M(\hat{\rho}^\Gamma)) \qquad (25)$$

The simulation of conditional correlation and the PT matrix determinant using the Schrödinger picture yields the same result. The anti-Stokes field losses do not change the sign of the determinant. This means that even for lower anti-Stokes field readout efficiency and higher mode mismatch, the relation between the tensor function and *MaxN* (Table1) is still valid. Figure 4 also retain their validity. This conclusion agrees with the previous remark that noise during anti-Stokes field readout doesn't necessarily weaken the entanglement test. Since the full entanglement criterion can be phrased in terms of all principle minor subdeterminants, the same conclusion applies to the full entanglement test, including the one using eigenvalues.

## 5. Quadrature measurements and Gaussian-state entanglement

Above we found that the criteria using moment measurement for number-state entanglement is sensitive to practical detection factors. Criteria using quadrature variance might be more powerful to verify the entanglement of the anti-Stokes states.

*5.1 Entanglement verification based on field-quadrature inequalities*

Balanced homodyne detection (BHD) is a robust technique for measuring field quadrature amplitudes. After the $n$ and $m$ photon-number measurements of Stokes fields after the 50:50 BS, the joint state of the two atomic-ensemble collective modes, in $q$ quadrature representation, is [10]

$$\Psi_{A,B} = \phi_n(\frac{q_B + q_A}{\sqrt{2}})\phi_m(\frac{q_B - q_A}{\sqrt{2}}) \qquad (26)$$

In which, $\phi_n(\bullet)$ is the $n$-th order Hermite-Gaussian function (harmonic oscillator energy eigenfunction). We assume that we are in the ideal situation: the measured Stokes state reflects the real atomic state, and this atomic state can be perfectly read out onto anti-Stokes fields and measured by noiseless detectors. The anti-Stokes field quadrature variances are calculated to compare with entanglement criteria developed in [2]. As shown there, a state is entangled if (but not only if)

$$\text{Var}(q_B + q_A) + \text{Var}(p_B - p_A) < 2 \qquad (27)$$

By changing variables and using properties of Hermite polynomials, we get for the variance, $\text{Var}(q_B + q_A) = 2n + 1$. The wave function in the $p$ quadrature representation can be obtained by taking the Fourier transform of Eq. (26), which gives

$$\Psi = \sum_{i=0}^{n+m} B_i^{n,m} \phi_i(p_A)\phi_{n+m-i}(p_B) = \phi_n(\frac{p_B + p_A}{\sqrt{2}})\phi_m(\frac{p_B - p_A}{\sqrt{2}}) \qquad (28)$$

By the same procedure, we get $\text{Var}(p_B - p_A) = 2m + 1$. Thus the total variance is

$$\text{Var}(q_B + q_A) + \text{Var}(p_B - p_A) = 2(n + m + 1) \qquad (29)$$

Unfortunately, this means that this entanglement, created by direct counting of Stokes photon numbers, can't be verified by BHD of the anti-Stokes fields and the entanglement criteria used here.

*5.2 Gaussian state quadrature variable entanglement generation and verification*

Instead of direct detection, if we measure the $p$ quadrature and the $q$ quadrature, respectively, of the two immerging Stokes fields after the BS, and substitute the observed values P and Q, then the conditioned wave function of the anti-Stokes fields (still assuming ideal condition) is [10],

$$\psi(q_A, q_B) = \exp\{-\frac{s(q_A + q_B)^2}{4} - \frac{(q_A - q_B)^2}{4s} + i\frac{2\sqrt{2}\mu\nu P(q_A - q_B)}{s} + 2\mu\nu\sqrt{2}Q(q_A + q_B)\} \qquad (30)$$

where $q_A$, $q_B$ are $q$ quadratures of anti-Stokes fields from cells A and B, $\mu = \cosh\alpha$, $\nu = \sinh\alpha$ and $s = \mu^2 + \nu^2$. This Gaussian state is an entangled state. The $q$ quadrature variance for the state in Eq. (30) equals $Var(q_B + q_A) = s^{-1}$. In order to get the $p$ quadrature variance, we need to take the Fourier transform with respect to $q_A$, $q_B$, then we get the $p$ quadrature wavefunction,

$$\psi(p_A, p_B) = \exp\{-\frac{(p_A + p_B - c_1)^2}{4s} - \frac{s(p_A - p_B + c_2)^2}{4}\} \qquad (31)$$

with $c_1$, $c_2$ complex constant numbers related to the squeezing parameter and measured value of P and Q. It is straightforward to get $\text{Var}(p_B - p_A) = s^{-1}$ and total variance,

$$\text{Var}(q_B + q_A) + \text{Var}(p_B - p_A) = 2s^{-1} \leq 2 \tag{32}$$

The quadature measurement by BHD is prone to a variety of inefficiencies, of which two main sources are the optical losses and the detector electronics noise. The optical losses are due to absorption in the beam paths, non-unitary efficiency of the detector, and imperfect mode matching between the signal and local oscillator. The effect of optical losses in the quantum state reconstruction is well understood, and there are ways of both its quantitative evaluation and compensation [20, 33]. The effect of electronic noise is recently shown to be equivalent to that of optical losses by Appel *et al.* [34] in their quadrature squeezing experiment,

$$\eta_{en} = \frac{\alpha^2}{\alpha^2 + \sigma^2} = 1 - S^{-1} \tag{33}$$

where $\alpha$ is the coefficient relating detector signal V to the field quadrature Q or P in the formula $V = \alpha Q$, $\sigma$ is the standard deviation of detector electronic noise, $S$ is the ratio between mean square noise of the vacuum state and the mean square electronic noise,

$$S = \frac{\alpha^2 + \sigma^2}{\sigma^2} \tag{34}$$

As evidenced by this equation, low electronic noise and strong signal will be helpful to improve the equivalent efficiency.

For the anti-Stokes fields from cell A and B undergoing an optical loss of $\eta_A = \eta_A = \eta$, the q-quadrature variance reduces by a factor of loss ratio, but gains excess vacuum noise of amount $1 - \eta$ [34],

$$Var((q_A + q_B)_\eta) = \eta \cdot Var(q_A + q_B) + (1 - \eta) \tag{35}$$

where $\eta$ is the overall effective loss efficiency considering both optical losses and detector electronic noise. From Eq. (35), if the q-quadrature variance in ideal situation is less than one, the q-quadrature variance in the loss situation will be still smaller than one, unless $\eta$ is 0. Thus a reasonable $\eta$ is sufficient to observe the violation of Eq. (32) in the Gaussian state entanglement. The overall effective loss efficiency $\eta$ of 51% is reported in the quadrature squeezing experiment [34].

## 6. Conclusions

In the number state entanglement scheme, the excitation number involved in the entanglement has great impact on the complexity of the verification scheme. In the microscopic regime, such as single-excitation entanglement [6], the photon correlation measurement by avalanche photo diode in Geiger mode is sufficient to verify the entanglement. In the mesoscopic region, for excitation level between 1 and 13, the visible light photon counter (VLPC) developed at Fermilab for a particle tracking system was reported to resolve up to seven photons [35]. In this situation, the simplest tensor function $\hat{f}_{1,2} = (1, \hat{a}^2 \hat{b}^2)$ is sufficient to reveal the entanglement and the corresponding matrix elements can be obtained through a relatively simple eight-port detection scheme [32]. As the excitation number goes above 13, we have to use the tensor function with higher order and more terms. The matrix elements can only be obtained through joint balanced homodyne detection [27, 29], or otherwise be obtained through more precise and also more complicated joint balanced homodyne correlation measurement [30, 31]. Currently BHC is only a theoretical proposal and no BHC experiment has been reported yet. As the excitation number gets bigger (say 14-400) the difficulty of detecting entanglement grows. In a summary, the larger the excitation number in the entanglement is, the more complicated the entanglement verification and generation scheme is. In the macroscopic limit (for example, millions of excitations), it presumably becomes impossible to verify the entanglement. One goal of this paper is to discover what the bounds of microscopic, macroscopic and mesoscopic are.

Using direct Stokes-photon detection for entanglement creation, the conditional anti-correlation of anti-Stokes photon numbers from two cells is found to be well preserved under various deviations from the ideal situation, such as the anti-Stokes field losses, Stokes field detector efficiency and additive thermal noise. But verification of entanglement, using moment measurements, is sensitive to these practical factors. There are two main reasons. First, current technology does not allow us to generate a pure number-state entanglement. Second, we might not have yet found the optimum criterion for entanglement verification. The extension of the SV approach provides only a sufficient—not necessary—condition for entanglement. It depends on the state to be measured.

We found that balanced homodyne detection can be used successfully for Gaussian state entanglement generation and verification. It is not required to perform full quantum state tomography; rather measuring certain quadrature variances is sufficient. The optical losses and detector noise in the verification stage don't have stringent effect on observing the violation of entanglement inequality.

## 7. Acknowledgement

This work has been done under NSF support with grants number PHY-0140370 and PHY-0456974.